\documentclass[final]{agujournal2019}
\usepackage{url} 

\usepackage{amsmath}
\usepackage{graphicx}
\usepackage{float}
\usepackage{diagbox}
\usepackage{slashbox}
\graphicspath{ {./Figures/} }
\usepackage{siunitx}
\usepackage{subcaption}
\usepackage{caption}
\usepackage{soul,color}
\usepackage{bm}

\journalname{JGR: Earth Surface}

\begin{document}

\title{The effect of grain size on erosion and entrainment in dry granular flows}

\authors{Eranga Dulanjalee\affil{1},
       Fran\c{c}ois Guillard\affil{1},
       James Baker\affil{1},
       Benjy Marks\affil{1}}

\affiliation{1}{School of Civil Engineering, The University of Sydney, Sydney, Australia}
\correspondingauthor{Benjy Marks}{benjy.marks@sydney.edu.au}

\begin{keypoints}
  \item The grain size of both the flowing and bed materials significantly influence erosion and entrainment in geophysical granular flows
  \item A new technique for measuring erosion is proposed which uses dynamic X-ray radiography to measure the grain size within the flowing material
  \item The measured erosion rate strongly depends on the measurement technique, with previous techniques giving inconsistent results
 
\end{keypoints}


\begin{abstract}

The entrainment of underlying erodible material by geophysical flows can significantly boost the flowing mass and increase the final deposition extent. The particle size of both the flowing material and the erodible substrate influence the entrainment mechanism and determine the overall flow dynamics. This paper examines these mechanisms experimentally by considering the flow of particles over an erodible bed using different particle size combinations for the incoming flow and the base layer in a laboratory-scale inclined flume. Dynamic X-ray radiography was used to capture the dynamics of the flow-erodible bed interface. The experiments found that the maximum downslope velocity depends on the ratio between the size of the flowing particles and the size of the bed particles, with higher ratios leading to faster velocities. Two techniques were then applied to estimate the evolving erosion depth: an established critical velocity method, and a novel particle-size-based method. Erosion rates were estimated from both of these methods. Interestingly, these two rates express different and contradictory conclusions. In the critical-velocity-based rate estimation, the normalized erosion rate increases with the flow to bed grain size ratio, whereas the erosion rates estimated from the particle-size-based approach find the opposite trend. We rationalise this discrepancy by considering the physical interpretation of both measurement methods, and provide insight into how future modelling can be performed to accommodate both of these complementary measures. This paper highlights how the erosion rate is entirely dependent on the method of estimating the erosion depth and the choice of measurement technique.

\end{abstract}

\section*{Plain Language Summary}

[Geophysical granular flows often cause severe destruction in mountainous terrains. The destructive potential of these flows is determined by the material and terrain properties and the dynamic interaction of the flow and the substrate. Erosion and entrainment of the substrate into the flow can cause significant volume expansion and increase the runout distance and the hazard to human safety. Here, we experimentally investigate these mechanisms with a particular interest in the effect of the relative particle size of the flowing and erodible substrate materials. A flow of particles over an erodible base was examined in a laboratory-scale inclined flume with different particle size combinations for the flow and the erodible substrate. The dynamics of the flow-erodible bed interface was captured with X-ray radiography. The evolving erosion depths were estimated from the well-established critical velocity based method and a novel particle size-based method. The erosion rates estimated from the two approaches gave contradictory conclusions. The study reveals that the erosion rate is entirely dependent on the measurement technique and the erosion depth evaluation method.]

\section{Introduction}

Geophysical granular flows such as landslides, debris flows and avalanches frequently threaten life, property and infrastructure in steep, hilly terrain. The destructive potential, size and flow velocity of these geophysical mass flows are strongly influenced by material rheology, terrain properties and the interaction mechanisms involved in the flow regime \cite{Mangeney:2011,Iverson2015}. The flow paths of the geophysical flows are usually covered with heterogeneous weathered geomaterials deposited in different geological events. During the mass movement process, the rapid flow interacts with the underlying surficial deposits and can incorporate these materials into the flowing system. 

\begin{figure}
    \centering
    \includegraphics[width=0.9\textwidth]{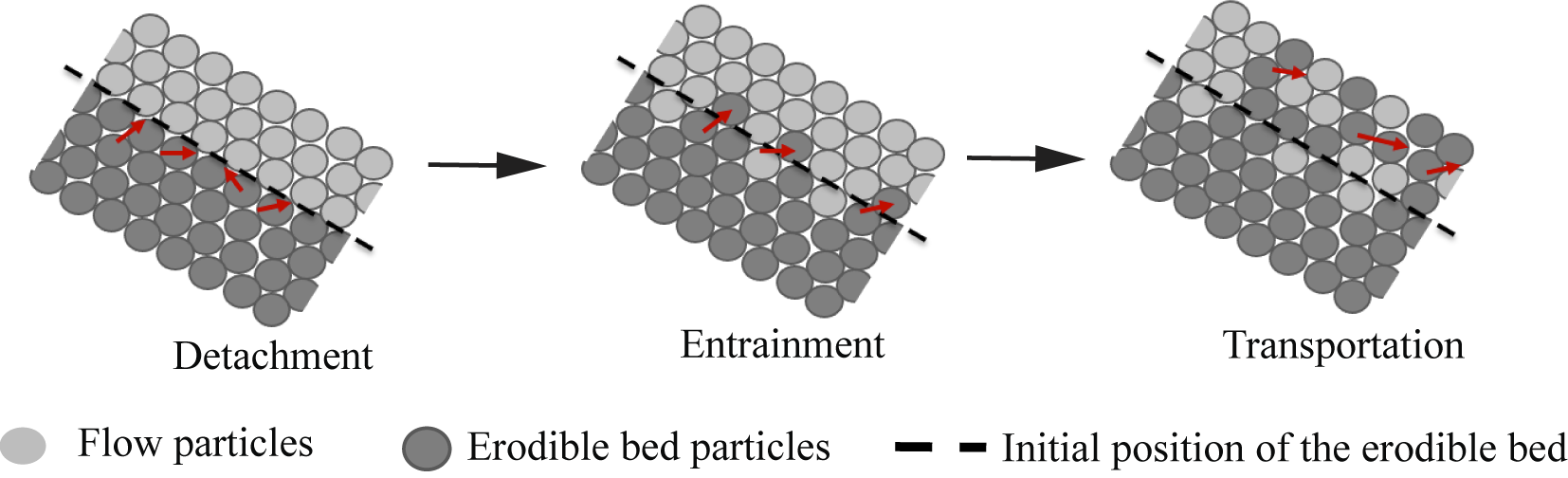}
    \caption{Schematic depiction of the grain-scale erosion process indicating detachment, entrainment and transportation.}
    \label{fig:1}
\end{figure}

Erosion is an important process in the dynamics of landslides, contributing to an increase in the flow volume and modifying the flow mobility by changing the material characteristics in the basal region of the flow mass \cite{Pirulli:2012}. The erosion process has been described as a combination of several successive mechanisms, namely detachment, entrainment and transportation, which are closely interrelated to each other \cite{Bracken:2015,Damme:2020} (see Figure~\ref{fig:1}). Entrainment is often interpreted as the key mechanism responsible for the erosion of the underlying bed material \cite{Mangeney:2011,Moberly2015,Iverson2015,Chao:2017}. According to field observations, the magnitude of the debris flow or the avalanche is rarely determined by its initial volume, as it often enlarges significantly due to the entrainment of materials along its path \cite{Hungr:2005,Farin:2014}. 

The erosion and entrainment mechanisms are interrelated to each other and it is crucial to distinguish the two mechanisms in experimental quantification. The rate of change of the bed elevation is the common measure of the basal erosion rate. \citeA{Chao:2017} provided a local scale definition for the entrainment rate as the material volume flux per unit area that enters into the flow. In early studies, the entrainment rate was correlated with parameters such as the difference between the angle of inclination and the repose angle of the materials \cite{Tai:2008}, excess boundary shear stress and basal flow velocity \cite{Iverson:2012}.

In order to understand the mechanics and conditions, erosion and entrainment experiments have been conducted at different scales using inclined flume configurations. The flow is typically captured with high-speed video cameras by observing the flow through a transparent side wall of the flume. The primary concern of those experiments has been to investigate the effect of different boundary conditions on erosion and to estimate the erosion rate \cite{Egashira:2001,Papa:2004,Mangeney:2010,Iverson:2011a,Farin:2014,Moberly2015,Lu:2016,Yunyun:2019,Ghasemi:2019}. 

A few studies have been conducted to investigate the impact of the flow and substrate grain size on entrainment. \citeA{Egashira:2001} and \citeA{Papa:2004} studied the erosion profiles of particle flow over different bed grain sizes. They concluded that the erosion rate monotonically decreases with increase of relative bed particle size. \citeA{Ghasemi:2019} concluded that bed erosion often increases with increasing the bed particle coarse fraction.

The use of laboratory scale flumes to model geophysical flows introduces several scaling issues \cite{Iverson:2010}. In addition, the presence of sidewalls means that the flow is strongly perturbed in these regions \cite{jop2005crucial} and that measurements made there are not representative of the bulk flow.
To overcome this limitation, many technologies have been applied to investigate the internal micro-mechanics of granular media including Magnetic Resonance Imaging (MRI) \cite{Stannarius:2017,Nakagawa:1993}, Positron Emission Particle Tracking (PEPT) \cite{Buffler:2018,Parker:1997}, X-ray computed tomography (X-ray CT) \cite{ando2012grain,Karatza:2019} and X-ray radiography \cite{Guillard2017,Baker:2018,Dulanjalee:20,Segolene:2021}. Each technique has its own advantages depending on the behavior of the flow regime and the focus of the investigation. Of these, X-ray radiography is a non-destructive investigative tool which has been developed to investigate internal micro-mechanical aspects of materials and multi-phase flows \cite{Heindel:2011}. X-ray radiography is a useful tool in granular flow research that is capable of capturing the internal evolving velocity field, particle size and orientation fields \cite{Guillard2017} as well as measuring the grain size fractions of evolving bidisperse materials \cite{Dulanjalee:20}. X-ray radiography captures the bulk flow behavior through the whole depth of the sample. Therefore, it minimises the effect of unrepresentative boundary layers that exist adjacent to the boundaries of experiments which are commonly observed with optical methods.

In a typical laboratory experiment, the only depth-dependent information comes from a high-speed camera which is positioned to observe through a transparent side wall. To approximate the erosion depth with such data, a `critical velocity method' \cite{Moberly2015} is commonly used to infer the position of the interface between the flowing material and the erodible bed using an arbitrary velocity cutoff in the downslope velocity profile. Alternative methods are available when further information on the grain types is measurable experimentally. In some reported experiments, different coloured particles were used for the flowing and erodible materials, such that the interaction between the materials could be measured using colour on the recorded video as a proxy \cite{Ghasemi:2019}. In this work, we have access to radiographs of the material over time, which can be used to infer the relative concentration of large and small particles in bidisperse granular mixtures directly \cite{Dulanjalee:20}. This `bidisperse calibration method' is a novel approach that relies on the ability of X-rays to pass through the medium, and characterise the micro-structure. By comparing the spatial length scale of the fluctuations of the density field, and with knowledge of the grain sizes, it is possible to estimate the local concentration of the two materials in a patch of an image. 

The purpose of this paper is to experimentally investigate the effect of grain size on the erosion mechanism using small-scale laboratory experiments. Specifically, we focus on different grain size combinations in the flowing and erodible bed materials on the erosion rate, which has not been systematically studied in previous work. In geophysical flows, the flowing material typically is not the same as the eroding material, and their particle sizes can vary widely due to heterogeneity of the in-situ material, as well as particle segregation and crushing during the flow \cite{johnson2012grain}. Additionally, the use of X-ray radiography as an imaging methodology allows us firstly to avoid the limitations in data capturing only near sidewalls, and secondly to enable measurement of the grain size within the flowing material.

\section{Materials and methods}

The main focus of this paper is the experimental investigation of the effect of grain size on the erosion mechanism during dry granular flow over an erodible bed. To investigate this experimentally, a laboratory-scale inclined flume setup was designed and the evolving flow field and grain size distribution was captured with dynamic X-ray radiography. 

\subsection{Experimental configuration}

\begin{figure}
    \centering
    \includegraphics[width=0.5\textwidth]{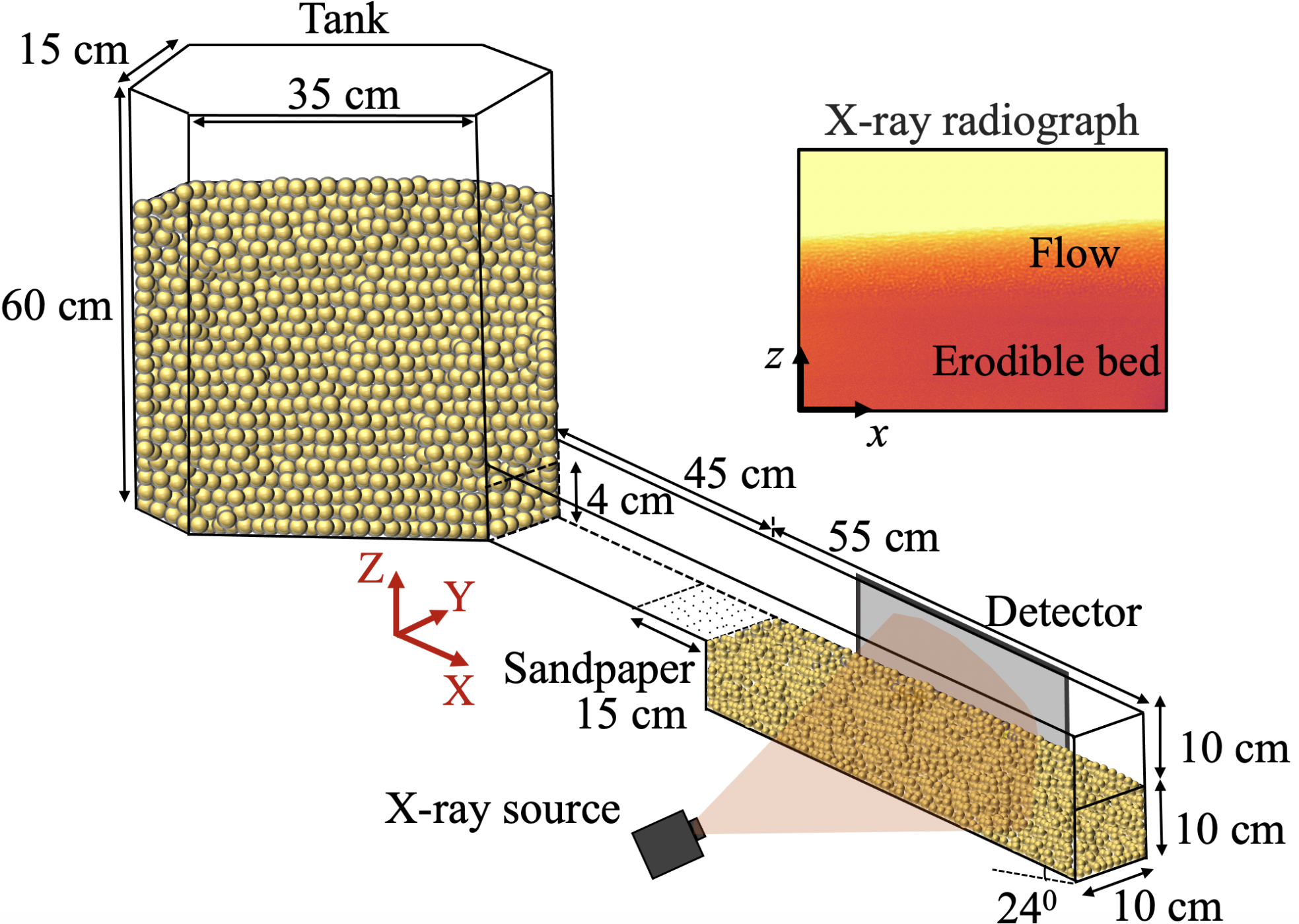}
    \caption[]{Schematic representation of the experimental setup with the X-ray arrangement. The X-ray radiograph (XZ direction) shows a view of the flow of 3~mm glass beads over 1~mm erodible bed.}
    \label{fig:2}
 \end{figure}

The experimental setup (as shown in Figure~\ref{fig:2}) is made of acrylic sheets and consists of a 1~m long flume with side walls spaced 10~cm apart. The flume is connected to a holding tank at the upstream end via an automated gate which is attached to a linear stage that raises the gate in the $Z$ direction. The material flow rate released from the tank is controlled by adjusting the final height of the automated gate. The flume bed consists of two sections. After particles exit the holding tank, they pass over a rigid, smooth acrylic base for a distance of 30~cm. For the last 15~cm of this section, sand paper of grit size 180 is glued to the rigid bottom to increase friction. The remaining section is composed of a separate chamber lowered 10~cm down from the initial flume bottom to hold the erodible layer (dimensions are as shown in Figure~\ref{fig:2}). All of the experiments reported here were conducted at a flume inclination angle of 24$^{\circ}$. We define a coordinate system in Figure~\ref{fig:2} with three mutually perpendicular orientations, $X$ being in the downslope direction, $Z$ being normal to the slope and $Y$ being the cross slope direction. $x$ and $z$ are the respective internal reference coordinates of the detector panel, parallel to the $X$ and $Z$ directions.

\begin{figure}
    \centering
    \includegraphics[width=0.6\textwidth]{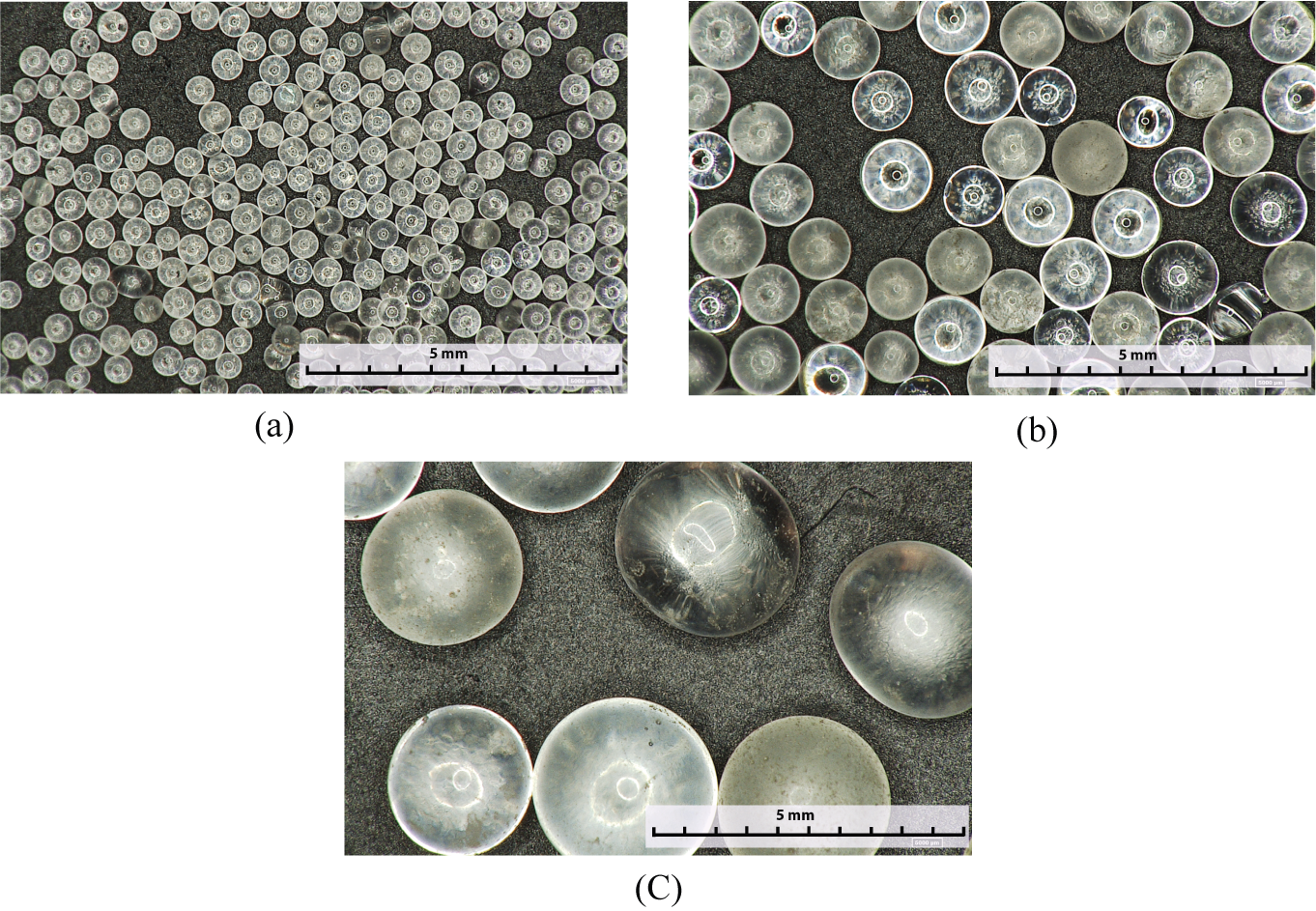}
    \caption[]{Microscopic view of the glass beads used for experiments (a) $d$=0.5~mm, (b) $d$=1~mm, (c) $d$ =3~mm.}
    \label{fig:3}
 \end{figure}

Soda lime glass beads of three sizes; nominally 0.5~mm, 1~mm and 3~mm diameter were chosen as the flowing and the erodible bed materials. The glass bead characteristics are shown in Table \ref{table:1} and microscope images of their morphology are shown in Figure~\ref{fig:3}. The typical diameter of the flowing and the bed materials are referred to as $d_{Flow}$ and $d_{Bed}$ respectively. The static repose angles were estimated using the hollow cylinder method \cite{Liu2011}. The values of the bulk densities were estimated by measuring the mass of a known volume (glass bead volume of a 500~mL graduated beaker)\cite{Jun2015}.

As shown in Table \ref{table:2}, eight experiments were conducted with different grain size combinations of the flowing and the erodible bed. Initially, the tank behind the closed gate was filled with the flow material to a height of 50~cm from the bottom of the tank. The erodible bed chamber was filled to the flume level (10~cm thickness) by pouring the bed material with a hand spade under gravity to form a loose bed. The surface of the bed was levelled with a piece of cardboard to maintain an even flow surface and uniform thickness. The experimental procedure attempted to conduct all experiments in similar flow conditions. Therefore, the experiments were conducted with the same gate height and gate opening speed. For all experiments reported here the gate was opened to a height of 40~mm, at an opening speed of 10~mm/s. As the gate opened, the material flow was initiated and the flowing mass was transported along the flume and over the erodible bed area. The materials that left the flume were collected in a bin at the end of the flume.
 
\begin{table}
    \begin{center}
        \caption{Characteristics of the glass beads used in the experiment.\label{table:1}}
        \begin{tabular}{ccc}
            \hline Grain size & Repose angle & Bulk density \\ 
                   (mm) & ($\pm$~1\textsuperscript{0}) & (kg/m$^{3}$) \\ 
            \hline
            0.5 & 20  & 1800 $\pm$ 12  \\ 
            \hline
            1 & 22  & 1712 $\pm$ 14  \\ 
             \hline
            3 & 24 & 1622 $\pm$ 15  \\
            \hline
        \end{tabular}
    \end{center}
\end{table}

\begin{table}
    \begin{center}
        \caption{Grain size combinations of flowing and erodible bed materials in each experiment (experiments labelled A-H).\label{table:2}}
        \begin{tabular}{c*{3}{c}}\hline
            \diagbox[innerwidth=35mm,height=3\line]{(mm)\\Bed grain size}{Flow grain size\\(mm)}
            &\makebox[2em]{0.5}&\makebox[2em]{1.0}&\makebox[2em]{3.0}\\\hline
            0.5 & - & A & B\\\hline
            1.0 & C & D& E\\\hline
            3.0 & F & G & H\\\hline
        \end{tabular}
    \end{center}
\end{table}
 
The motion of the flowing glass beads over the erodible bed was captured with dynamic X-ray radiography. The X-ray source was aligned with the flow to capture the flow field from one side of the flume wall (X-ray beam pointing in the Y direction, and imaging in the XZ plane)(see Figure~\ref{fig:2}). The X-ray source was aligned to capture the centre of the middle region of the erodible bed chamber. A Varian NDI-225-21 X-ray tube was used to emit radiation at a maximum energy of 180~keV and intensity of 8~mA. This radiation passed through the grains in the flume, and the transmitted radiation was recorded on a PaxScan 2520DX detector panel at a resolution of 768~px $\times$ 960~px and recording speed of 30 frames per second. The measured intensity of X-rays ($I$) at a location ${\bm x}$ on the detector panel can be approximated by the Beer-Lambert law as,

\begin{equation}
    I({\bm x}) = I_0 \exp\left(-\int \mu(l)~dl\right),
    \label{1}
\end{equation}

\noindent where $I_{0}$ is the intensity of the source and $\mu(l)$ is the mass attenuation coefficient for the X-ray radiation, at a location $l$ along the line of sight between the X-ray source and the pixel of interest. For materials of a single chemical composition, the mass attenuation coefficient is proportional to the density of the material \cite{Hubbell1982}. Therefore, the radiographs provide a local measurement of the density integrated along the ray path.

\subsection{Flow height measurement}

\begin{figure}
    \centering
    \includegraphics[width=0.8\textwidth]{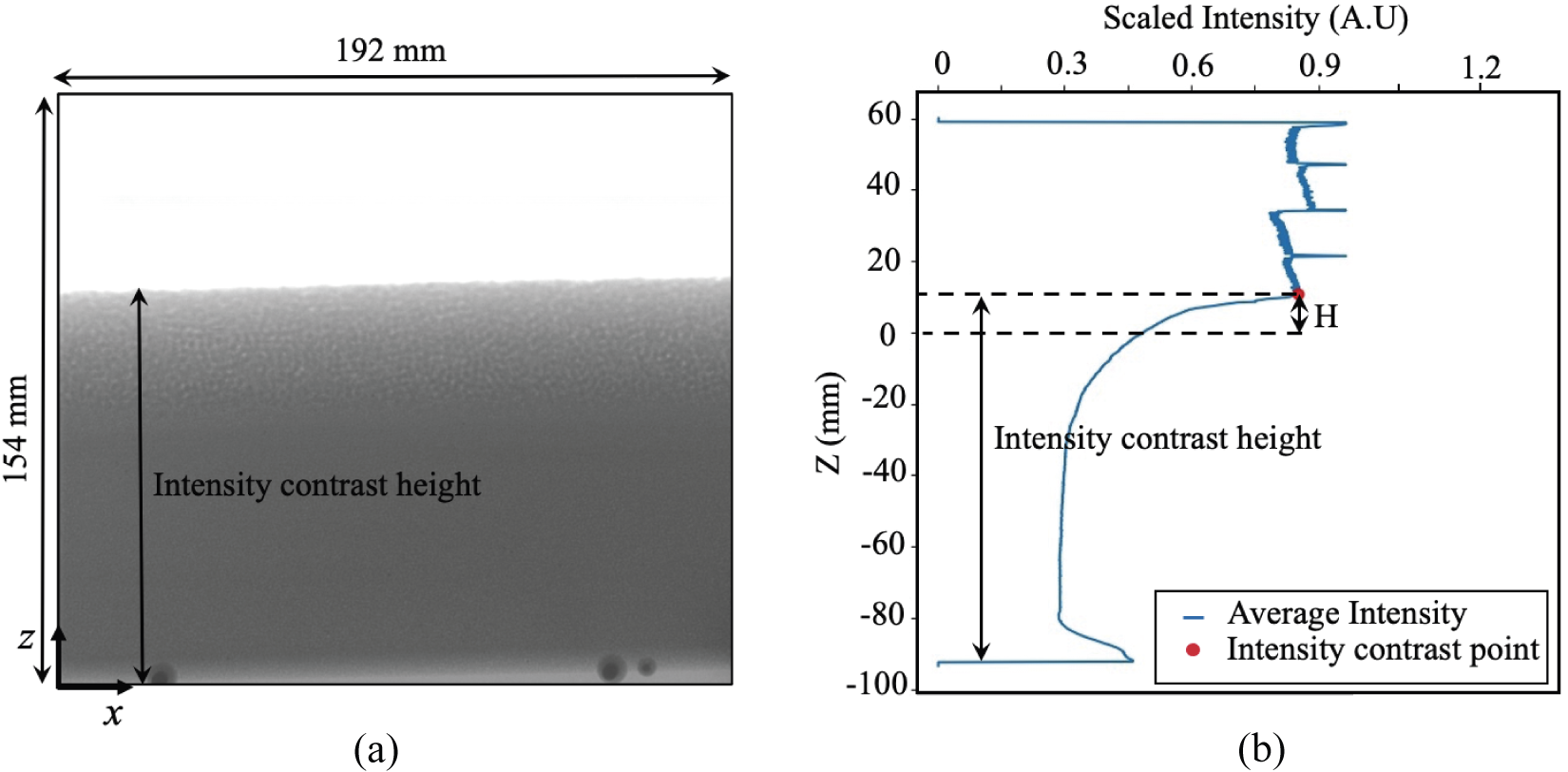}
    \caption[]{Intensity contrast height determination (Example from experiment $E$ at time t=3.08~s) (a) X-ray radiograph showing the intensity contrast between the granular materials and the particle-free region. (b) Average ($x$ direction) scaled intensity (in arbitrary units) variation with depth $Z$. The red dot indicates the detected free surface location. Here, the flow height is defined as the maximum material height from the initial position of the erodible bed ($H$). Z=0 is the initial position of the erodible bed.}
    \label{fig:4}
 \end{figure}

As shown in Figure~\ref{fig:4}(a) the radiographs exhibit a clear intensity contrast between the granular materials (dark regions on radiographs) and the particle-free zones (white regions). The interface between these two regions corresponds to the flow height at that particular time frame. To detect this interface, we averaged the X-ray intensity along the $x$ direction of the radiograph. The intensity of the X-ray beam is significantly lower when passing through the glass beads than through air (see Figure~\ref{fig:4}(a and b)). The interface between these two regions is delineated by a sharp change in intensity, which was used to identify the intensity contrast height.

\begin{figure}
    \centering
    \includegraphics[width=0.49\textwidth]{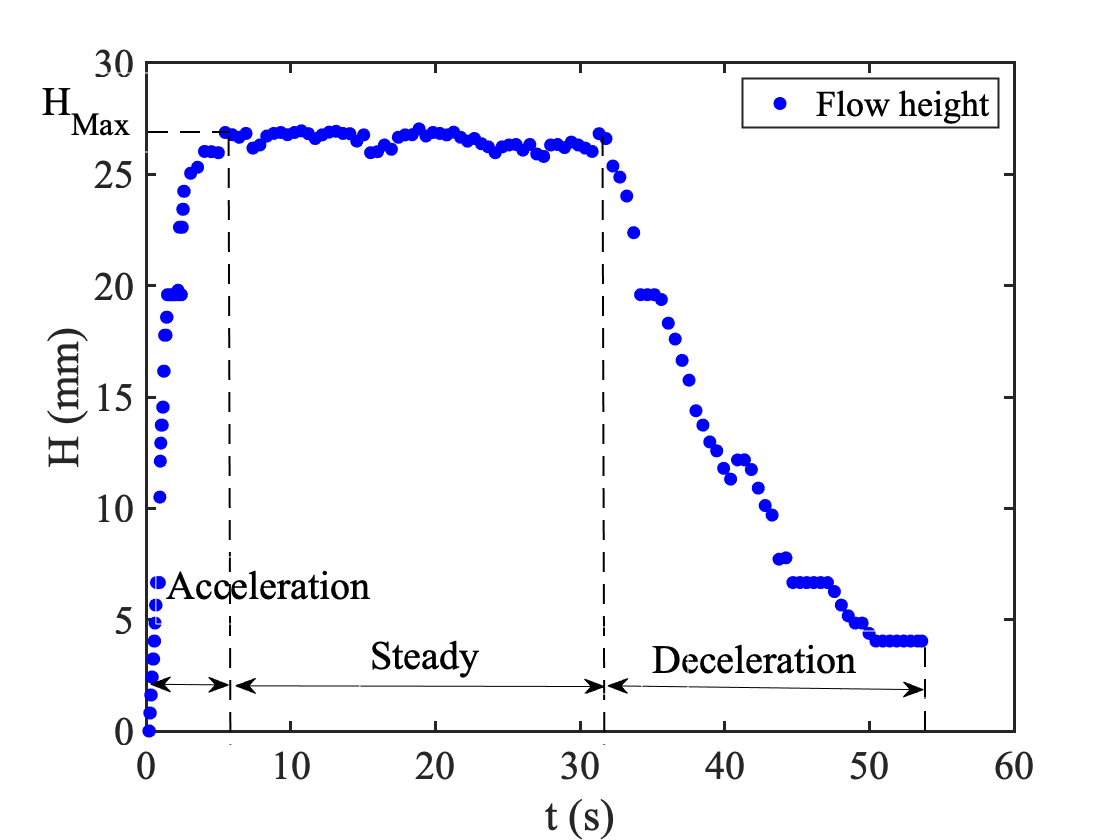}
    \caption[]{Variation of the flow height (Measured from the initial position of the erodible bed) with time of the experiment $E$.}
    \label{fig:5}
\end{figure}

When the gate is opened, the flowing material is suddenly released. The generated flow undergoes three distinct phases: acceleration, steady flow and deceleration (Figure~\ref{fig:5}). During the acceleration phase, the flow height gradually increases as material is released from the gate. At steady state the flow achieves a maximum uniform flow height $H_{Max}$. The decelerating phase occurs when the material in the tank empties.

\subsection{Flow velocity measurement}

Since granular materials are inhomogeneous, the radiographs depict the local density fluctuations of the flowing and the bed materials as they evolve in time. These series of evolving images were used to measure displacements over time using particle image velocimetry (PIV). PIV measures the velocity vector field between two consecutive images by applying cross-correlation analysis to deduce the most likely displacement in each selected image patch \cite{Adrian2011}. Here, the cross-correlation peak represents the modal displacement inside the measurement volume between two frames. This approach gives a measurement of two in-plane velocity components. The velocity estimated from X-ray radiography is more representative (when compared to PIV on more common optical images \cite{gollin2017performance}) of the motion far from boundaries, as it estimates the most probable velocity value along the beam.

The PIV tool PIVlab \cite{Thielicke_2014} is used to extract the velocity field from the radiographs. PIV analysis typically consists of three steps; image pre-processing, evaluation and post-processing. In pre-processing, contrast limited adaptive histogram equalization (CLAHE) was applied to normalise the images. Patches of 64$\times$64 pixels with 50~\% overlap were processed (approx 8$\times$8 mm) to find the typical displacement field. In post-processing, vector validation was performed by manually setting velocity limits to remove unphysical values. 

\begin{figure}
    \centering
    \includegraphics[width=0.8\textwidth]{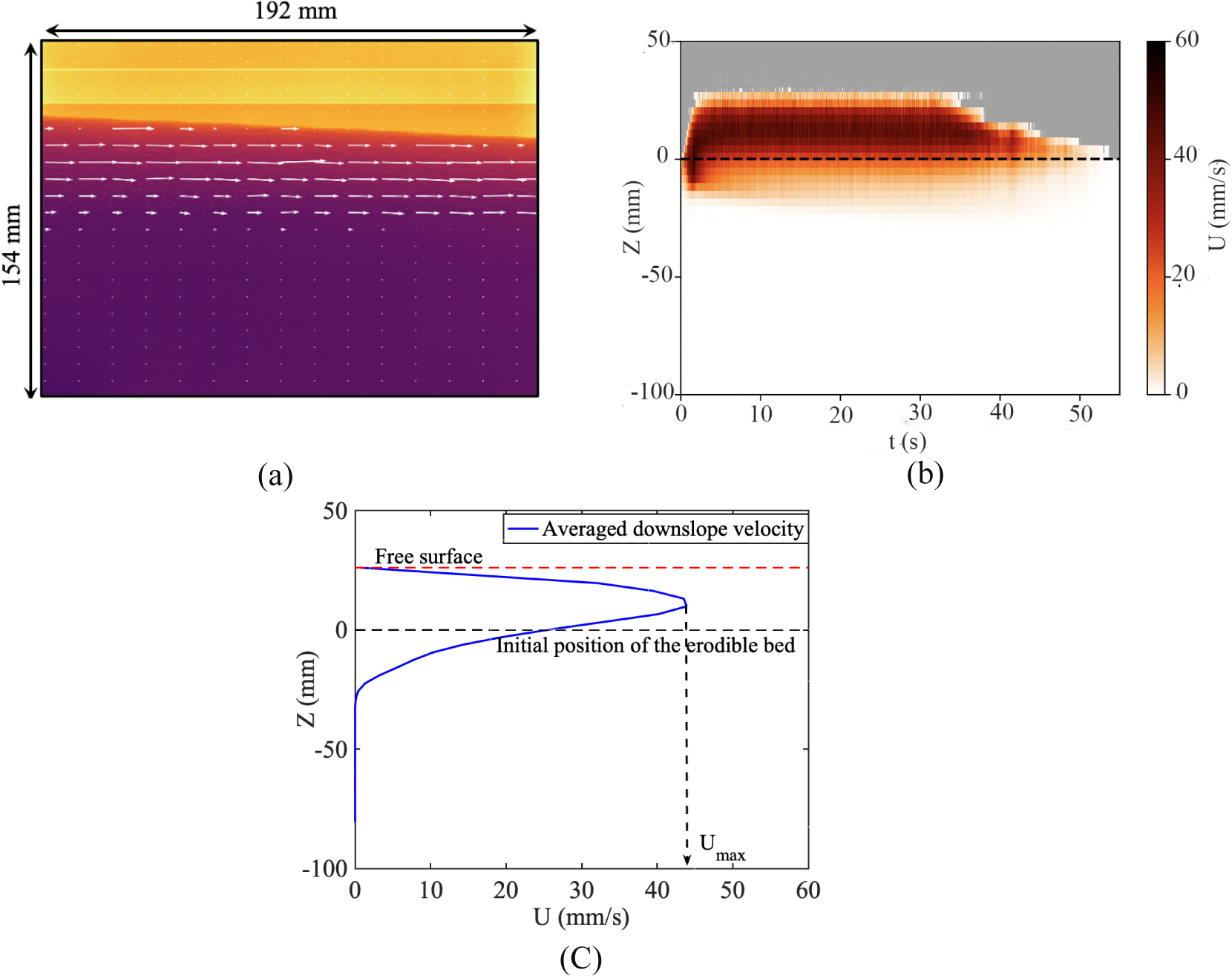}
    \caption[]{Flow velocity distribution (Example from experiment E). Depths are measured from the initial position of the erodible bed. (a) Instantaneous PIV velocity vectors superimposed with $xz$ radiograph at steady flow state. Note that many vectors have been omitted for clarity. (b) Spatio-temporal distribution of downslope velocity. Grey colour region indicates the area with no particle flow. (c) Averaged downslope velocity variation at steady flow state.}
    \label{fig:6}
\end{figure}

The recovered PIV velocity field superimposed over an $xz$ radiograph (Figure~\ref{fig:6}(a)) shows a typical downslope velocity field of the inclined flume. A static bottom layer, a dense flowing layer which extends from the static bottom layer to the maximum velocity level and an agitated fast flowing thin layer on top. The spatio-temporal distribution of the averaged downslope velocity (Figure~\ref{fig:6}(b)) gradually increases with time until it reaches a maximum velocity. This velocity phase exists for a certain time and velocity gradually decreases as the material in the tank empties. The temporally and spatially averaged downslope velocity at steady flow state (Figure~\ref{fig:6}(c)) decays exponentially with the distance from the top layer. The free surface level is estimated from the maximum flow height as shown in Figure~\ref{fig:5}. The velocity field appears to decrease between the maximum velocity and the free surface.This may indicate that the velocity of this thin layer is too fast to trace by PIV or that the flow density decreases to zero in this region, or both. 

\subsection{Erosion depth and erosion rate determination}

\subsubsection{Erosion depth}

\begin{figure}
    \centering
    \includegraphics[width=0.8\textwidth]{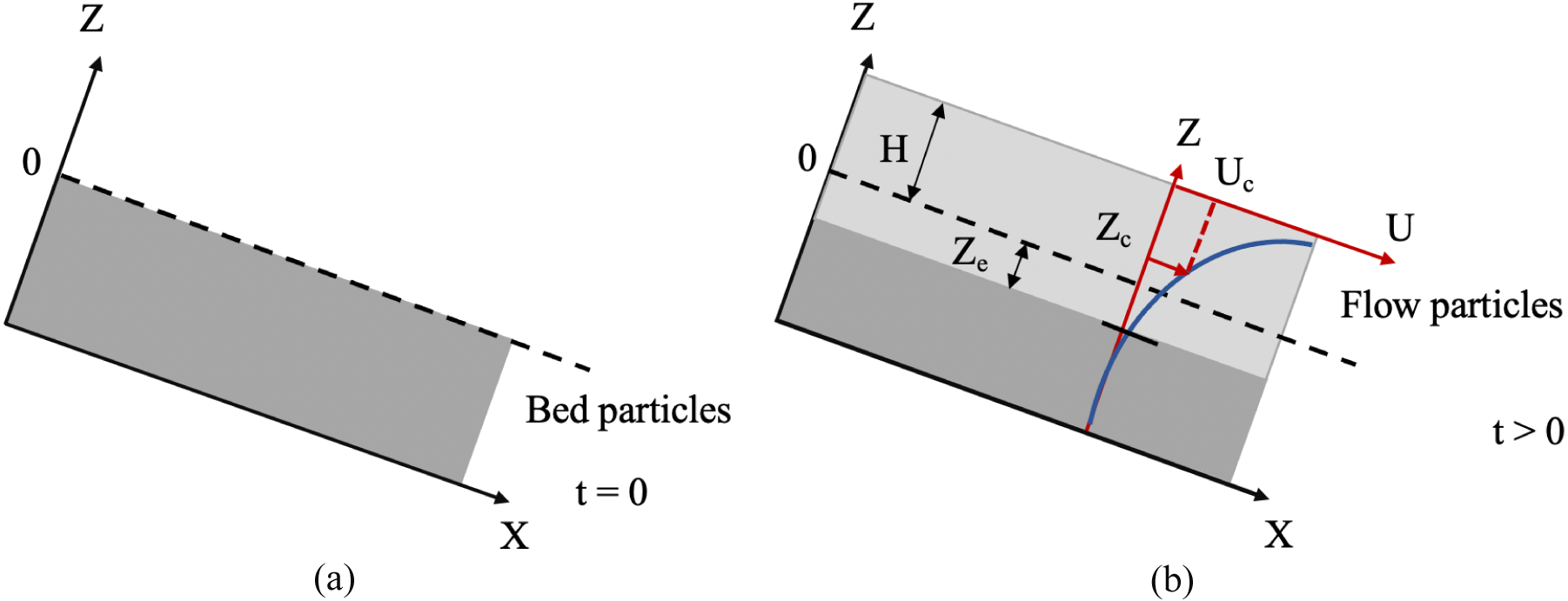}
    \caption[]{Schematic depiction of an eroding  boundary where $Z_e$ and $H$ are the erosion depth and the flow height above the initial position of the erosion bed respectively. $U_c$ is the arbitrarily defined critical velocity at which the flowing and erodible bed interface is positioned in the critical velocity method and $Z_c$ is the respective depth at the critical velocity.}
    \label{fig:7}
\end{figure}

The erosion depth is determined based on the position of the upper level of the erodible bed. The erosion depth ($Z_e$) is calculated as the elevation difference between the current and the initial position of the erosion bed (Figure~\ref{fig:7}). Therefore, a negative signed erosion depth value indicates erosion, and deposition is indicated with a positive sign. Two methods are used below to estimate this erosion depth.

\subsubsection{Critical velocity method}

As is typical for flows over erodible beds, the downslope velocity decreases exponentially near the flow-erodible bed interface (Figure~\ref{fig:6}(b)). The depth at which the downslope velocity becomes zero resembles the `motion' and `no-motion' interface (flow-erodible bed interface). In an exponentially decreasing velocity trend it is difficult to estimate the height at which the velocity reaches a value of zero. Therefore, a critical velocity value ($U_c$) of 0.001~ms$^{-1}$ was assumed as the velocity of the flow-erodible bed interface. This value was chosen arbitrarily, although the final results are not sensitive to small changes in this value. 

\subsubsection{Bidisperse calibration method}

\begin{figure*}
    \centering 
    \includegraphics[width=0.95\textwidth]{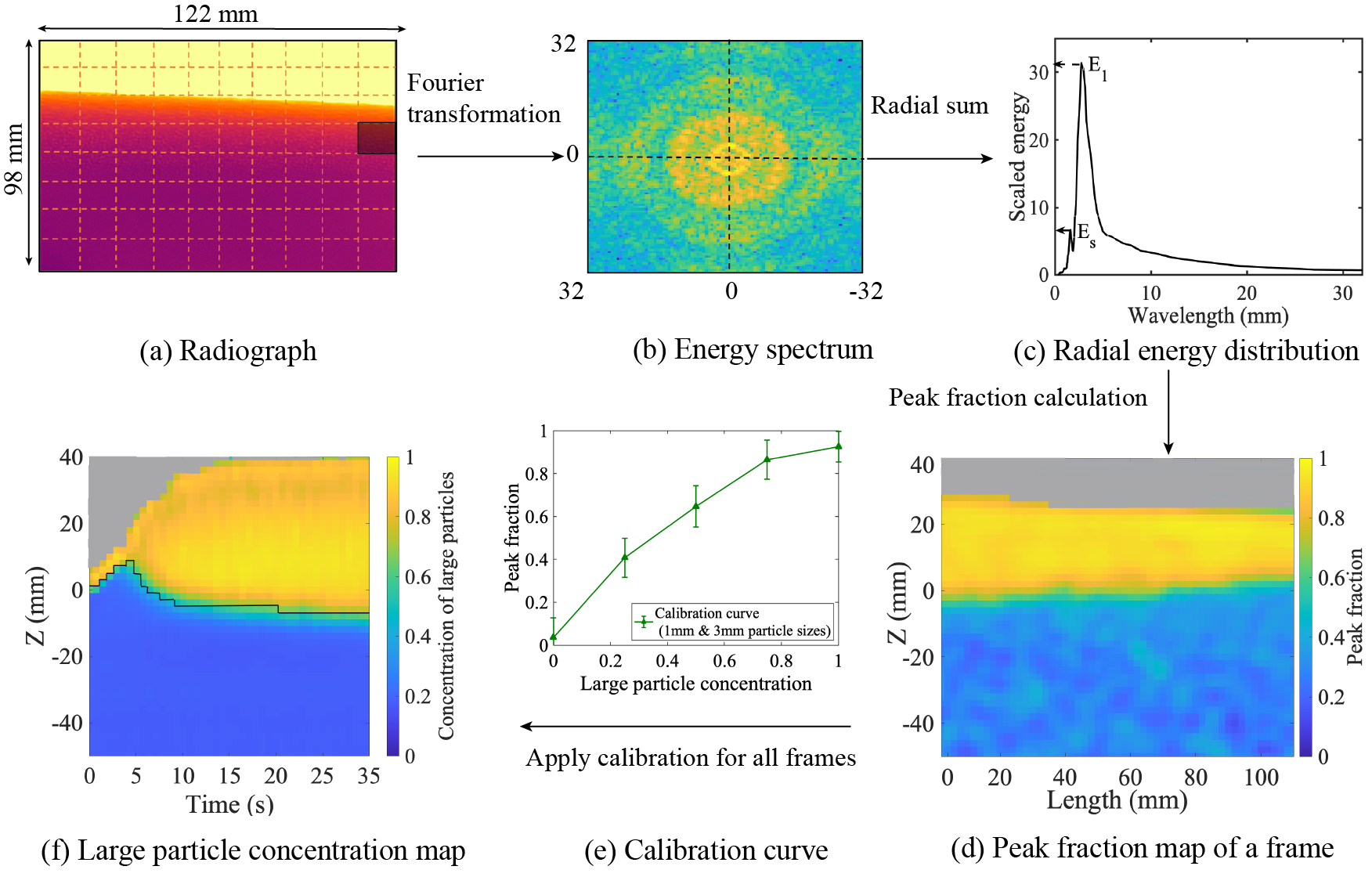}
    \caption{Determination method of bidisperse concentration (Example from experiment E and time t=5.0~s (a) initial radiograph indicating the patches of 64$\times$64 pixels (not to scale) (b) Energy spectrum of the patch (after multiplication by a circular hamming window and processed by two-dimensional Fourier transformation) (c) The scaled energy distribution as a function of the radius (d) Peak fraction map (e) Calibration curve of 1~mm and 3~mm particle combination. (f) Spatio-temporal distribution of concentration of large particles for the whole experiment. Black line is the $50\%$ concentration line of the bed grain size. Grey region is the region with no particle flow.}
    \label{fig:8}
\end{figure*}

In this work a new measure of the interfacial layer position is proposed and measured. We refer to this as the `bidisperse calibration' method, in contrast with the above `critical velocity' method. This bidisperse calibration technique has previously been developed to determine the relative volume fractions of the grain sizes in evolving bidisperse granular materials \cite{Dulanjalee:20}. The volume fraction is uniquely defined by the concentration of the large particles of the bidisperse granular system at a certain representative volume in space and time. The bidisperse calibration methodology is based on using Fourier transforms of X-ray radiographs (Figure~\ref{fig:8}(a)) to extract local measurements which evolve over time and can be related to the particle size distribution (Figure~\ref{fig:8}(b)). The technique measures the relative concentration of the two distinct species via a heuristic calibration parameter ``peak fraction''. This peak fraction is a measure of the relative height of two peaks in the orthoradially summed energy spectrum at the two spatial wavelengths of the particle sizes of the bidisperse mixture (Figure~\ref{fig:8}(c)). Three sets of calibration curves (one for each of the three bidisperse systems: 1\,mm-3\,mm combination, 0.5\,mm-1\,mm combination and 0.5\,mm-3\,mm combination) were generated using this approach to interpret the relationship of the concentration of the large particles in the bidisperse system and the measured peak fraction. Peak fraction maps were generated from the series of radiographs (Figure~\ref{fig:8}(d)). The peak fraction maps were converted to a concentration map based on the relevant calibration curve (Figure~\ref{fig:8}(e)) which was produced using the technique shown in \citeA{Dulanjalee:20}. The position of the erodible bed was assumed to exist at a concentration value of $50\%$ of the bed grain size, although the results are largely insensitive to changes of this value as the mixing zone is typically small (Figure~\ref{fig:8}(f)). 

Two sets of erosion profiles were generated from the erosion depth values estimated using both the critical velocity method and the bidisperse calibration method (see Supplementary Information Figure S7 for more information). The two methods lead to distinctly different erosion depth measurements. The critical velocity-based erosion profiles generally follow the flow phases, with a transient phase as the flows start, a steady phase, and deceleration at the end of the flow. Rapid erosion during the flow initiation is a common feature for experiments where the flowing particle size is greater than the bed particle size. The experiments where the flowing particle size is similar or smaller than the bed particle size exhibit rapid erosion and subsequent deposition during the flow initiation. During steady flow, both erosion (Experiments $C$, $E$, $F$ and $G$) and deposition (Experiments $A$, $B$, $D$ and $H$) conditions are observed. The deposition condition can be interpreted in two ways, corresponding either to the deposition of the eroded materials from upstream or a decrease of the flow velocity. Concentration-based erosion profiles exhibit a rapid increase of the bed level at during the flow initiation phase. Immediately after this increase, the erosion depths tend to decrease with time and reach a steady erosion depth during the later stage (deceleration phase) of the flow. The erosion depth profiles show three different trends of erosion namely, initial rapidly eroding profiles ($A$,$C$), shallow eroding profiles ($F$,$G$) and a combination of both ($B$,$E$). 

\subsubsection{Erosion rate}

The erosion rate ($\dot{e}$) is defined as the rate of change of the erosion depth, and can be approximated by 

\begin{equation}
   \dot{e} = \frac{{\Delta Z_e}}{{\Delta}{t}},
    \label{2}
\end{equation}

\noindent where $\Delta Z_e$ is the change in  erosion depth and ${\Delta}{t}$ is the time step between measurements (for the experiments reported here this is a constant of 1/30\,s).

\begin{figure}
    \centering
    \includegraphics[width=0.49\textwidth]{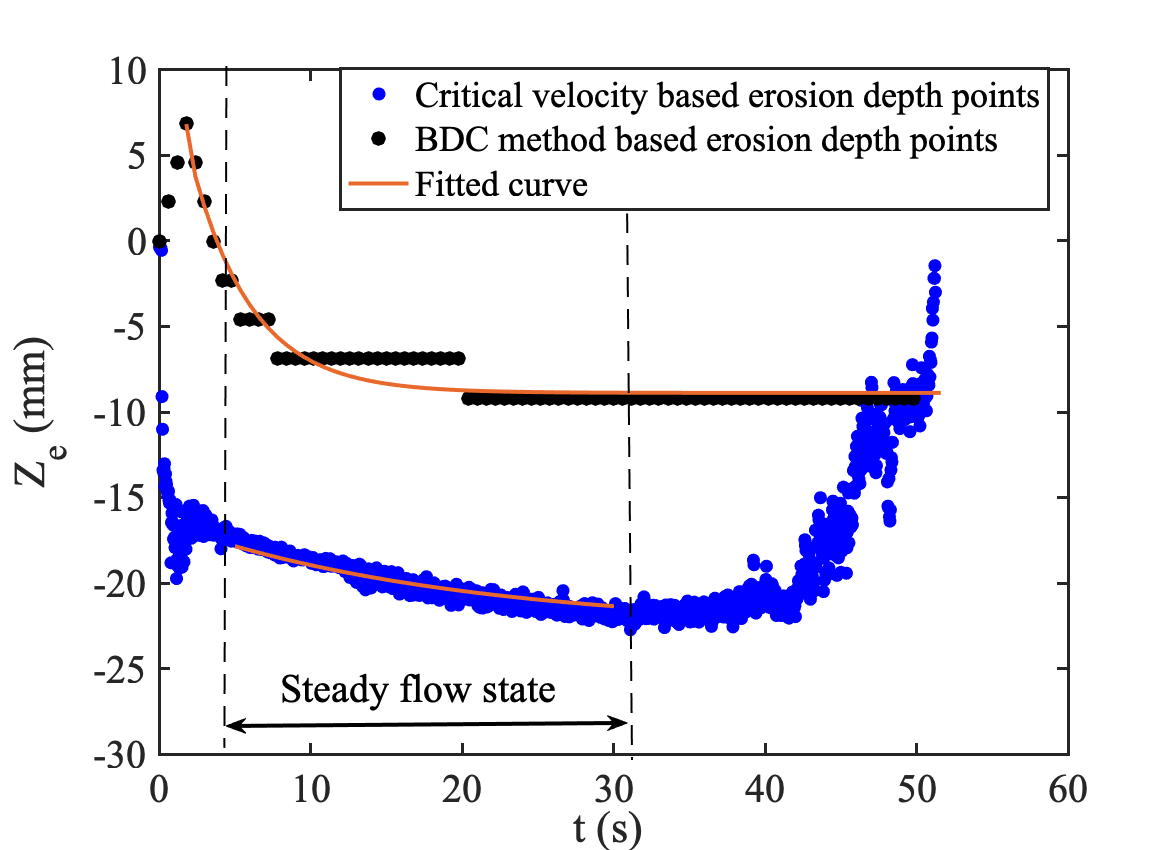}
    \caption[]{Erosion profiles generated from both critical velocity method and bidisperse calibration method (BDC). Example from experiment E.}
    \label{fig:9}
\end{figure}

The erosion rates were estimated by fitting an exponential curve to the erosion profiles. In the critical velocity approach, the erosion rate was estimated by fitting an exponential curve to the erosion profile segment during steady flow state, as during changes in flow behaviour (such as near the onset and cessation of flow) this measure is not representative of the erosion depth. For the bidisperse calibration method, an exponential curve was fitted from the maximum tip of the profile onwards (after the first bulbous head of the flow has passed, Figure~\ref{fig:9}). In both cases, the equation of the exponential curve fit is given by

\begin{equation}
   {Z_e}(t) = {Z_e^{max}}e^{-t/t_d} +c,
    \label{3}
\end{equation}

\noindent where ${Z_e^{max}}$ is the total erosion depth, $t_d$ the decay time and $c$ a constant. Reference erosion rates, which are representative of the whole time dependent behaviour were estimated from the fitted values of the constants as $\dot{e}_{ref}=-Z_e^{max}/t_d$. 

Erosion rates estimated from the exponential curve fitting are shown in Tables~\ref{table:3} and \ref{table:4}. In comparison with the bidisperse calibration method, the critical velocity method captures significantly smaller erosion rates. This discrepancy is linked to the fact that for the bidisperse calibration method the total erosion profile is considered for the rate estimation, whereas for the critical velocity approach only the steady flow state is considered (as shown in Figure~\ref{fig:9}). 

\begin{table}
    \begin{center}
        \caption{Erosion rate values calculated from the critical velocity method (mm\,s$^{-1}$).\label{table:3}}
        \begin{tabular}{c*{3}{c}}\hline
            \diagbox[innerwidth=35mm,height=3\line]{(mm)\\Bed grain size}{Flow grain size\\(mm)}
             &\makebox[2em]{0.5}&\makebox[2em]{1.0}&\makebox[2em]{3.0}\\\hline
             0.5 & - & 0.5 & 0.2\\\hline
             1.0 & -0.4 & 0.3 & -0.3\\\hline
             3.0 & -0.5 & -0.4 & 0.1\\\hline
        \end{tabular}
    \end{center}
\end{table}

\begin{table}
    \begin{center}
        \caption{Erosion rate values calculated from the bidisperse calibration method (mm\,s$^{-1}$).\label{table:4}}
        \begin{tabular}{c*{3}{c}}\hline
            \diagbox[innerwidth=35mm,height=3\line]{(mm)\\Bed grain size}{Flow grain size\\(mm)}
             &\makebox[2em]{0.5}&\makebox[2em]{1.0}&\makebox[2em]{3.0}\\\hline
             0.5 & - & -8.6 & -12.1\\\hline
             1.0 & -6.1 & - & -7.7\\\hline
             3.0 & -3.7 & -1.7 & -\\\hline
        \end{tabular}
    \end{center}
\end{table}

\section{Results and discussion}

\begin{figure}
    \centering
    \includegraphics[width=0.49\textwidth]{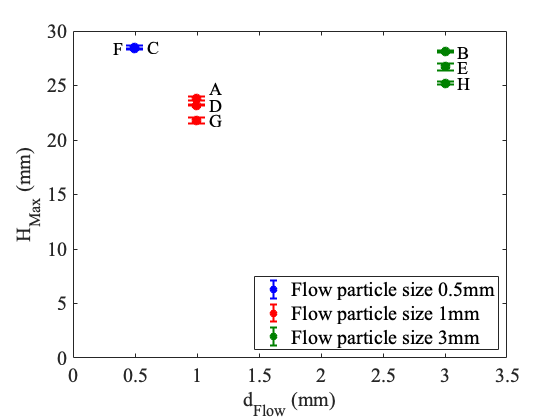}
    \caption[]{Average maximum flow height at steady flow state. Error bars represent one standard deviation of the measurement.}
    \label{fig:10}
\end{figure}

In Figure~\ref{fig:10} the average maximum flow height during the steady flow state is shown. It can be seen that experiments with identical flowing grain size have similar flow heights, and can be considered as representing similar flow conditions. Larger basal grain sizes tend to decrease the depth of the flowing layer.

\begin{figure}
    \centering 
    \includegraphics[width=0.5\textwidth]{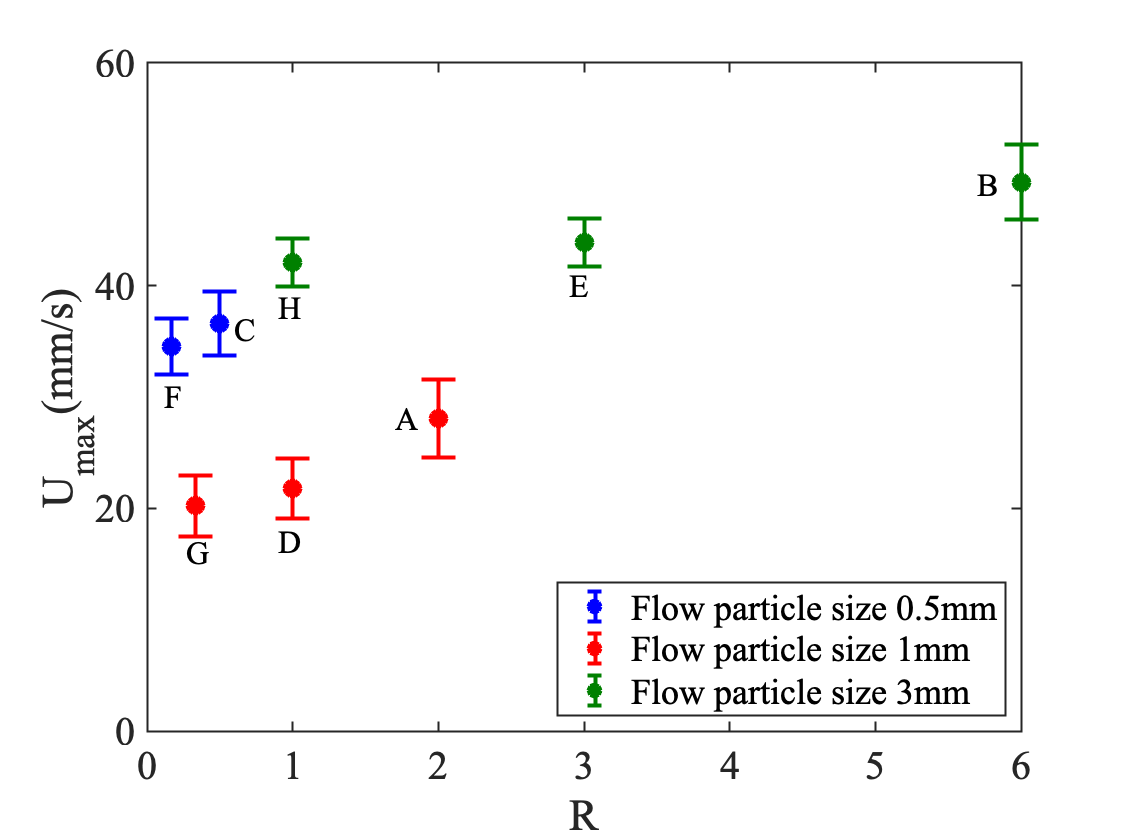}
    \caption{Variation of the maximum downslope velocity $U_{Max}$ with flow to bed particle size ratio $R$. Error bars represent one standard deviation of the measured value.}
    \label{fig:11}
\end{figure}

The relationship between the maximum average downslope velocity $(U_{Max})$ at the steady flow state and the flow to bed particle size ratio, $R=d_{Flow}/d_{Bed}$ is shown in Figure~\ref{fig:11}. Here, $d_{Flow}$ and $d_{Bed}$ are the diameters of the flow and bed materials respectively. As shown in Figure~\ref{fig:11}, $U_{max}$ increases as $R$ increases for a given flow particle size. 

\begin{figure*}
    \centering 
    \includegraphics[width=0.89\textwidth]{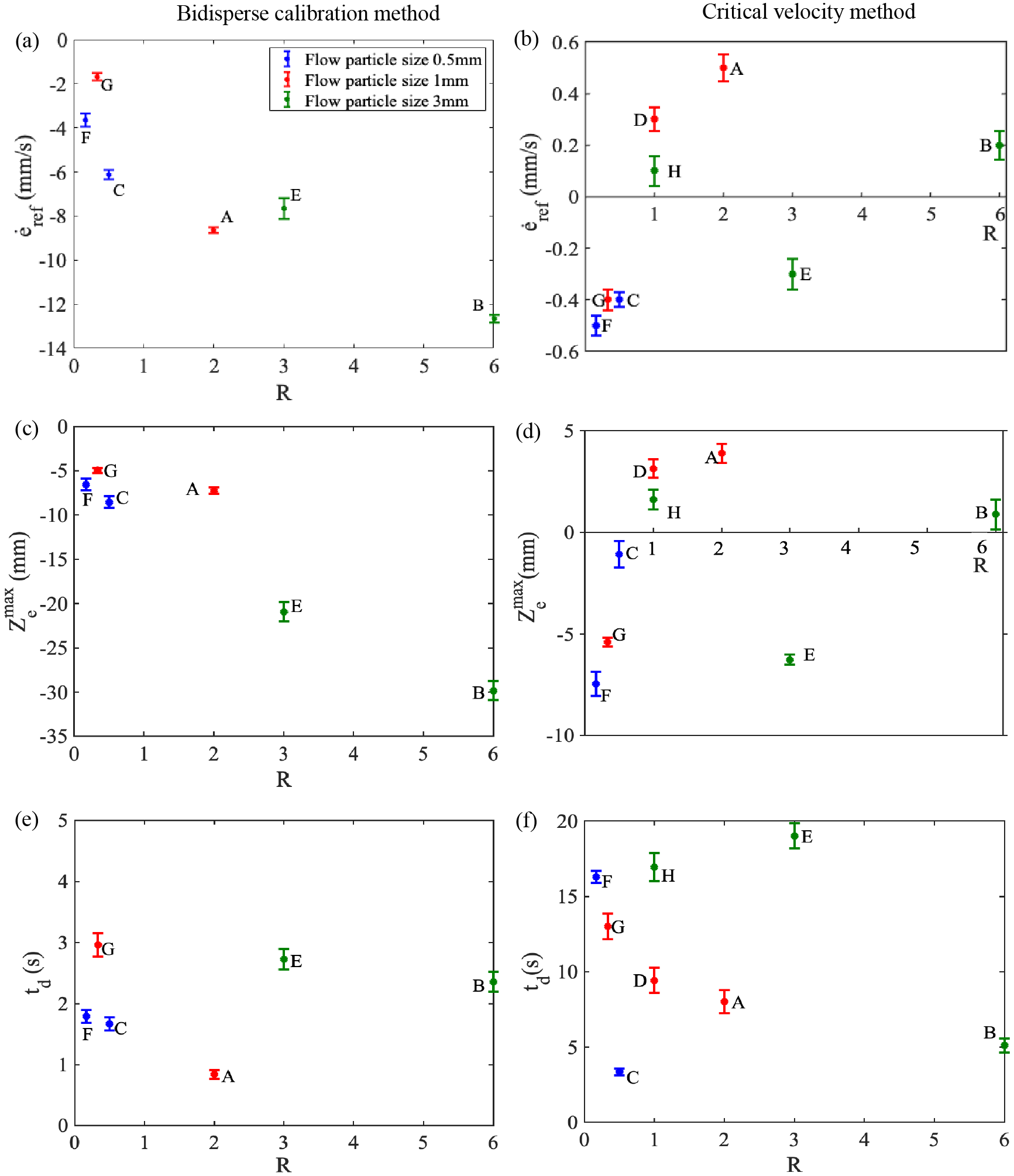}
    \caption{Comparison of the critical velocity and concentration based methods. (a), (b) Erosion rate. (c), (d) Maximum erosion depth. (e), (f) Decay time.}
    \label{fig:12}
\end{figure*}

A comparison of the two erosion methods concerning the reference erosion rate, maximum erosion depth, and the decay time is shown in Figure~\ref{fig:12}. The reference erosion rates estimated from the two methods exhibit opposite trends (see Figure~\ref{fig:12}(a) and (b)). For a given flow particle size, the erosion rates determined from the bidisperse calibration method increase with the flow to bed particle size ratio. Note that negative values indicate erosion. The reference erosion rate is a function of the maximum erosion depth and the decay time. In both methods the maximum erosion depth variation (Figure~\ref{fig:12}(c) and (d)) show a correlation with erosion rate. However, the decay time is relatively insensitive to the particle size ratio (Figure~\ref{fig:12}(e) and (f)). The decay time is the regulating factor of the erosion rate. This can be explained by an example comparison of the cases $A$ and $E$. Even though the experimental case $E$ has a higher maximum erosion depth than $A$, case $A$ exhibits a greater erosion rate. Here, the decay time regulates the erosion rate with greater decay time in $E$.

\begin{figure*}
    \centering 
    \includegraphics[width=0.9\textwidth]{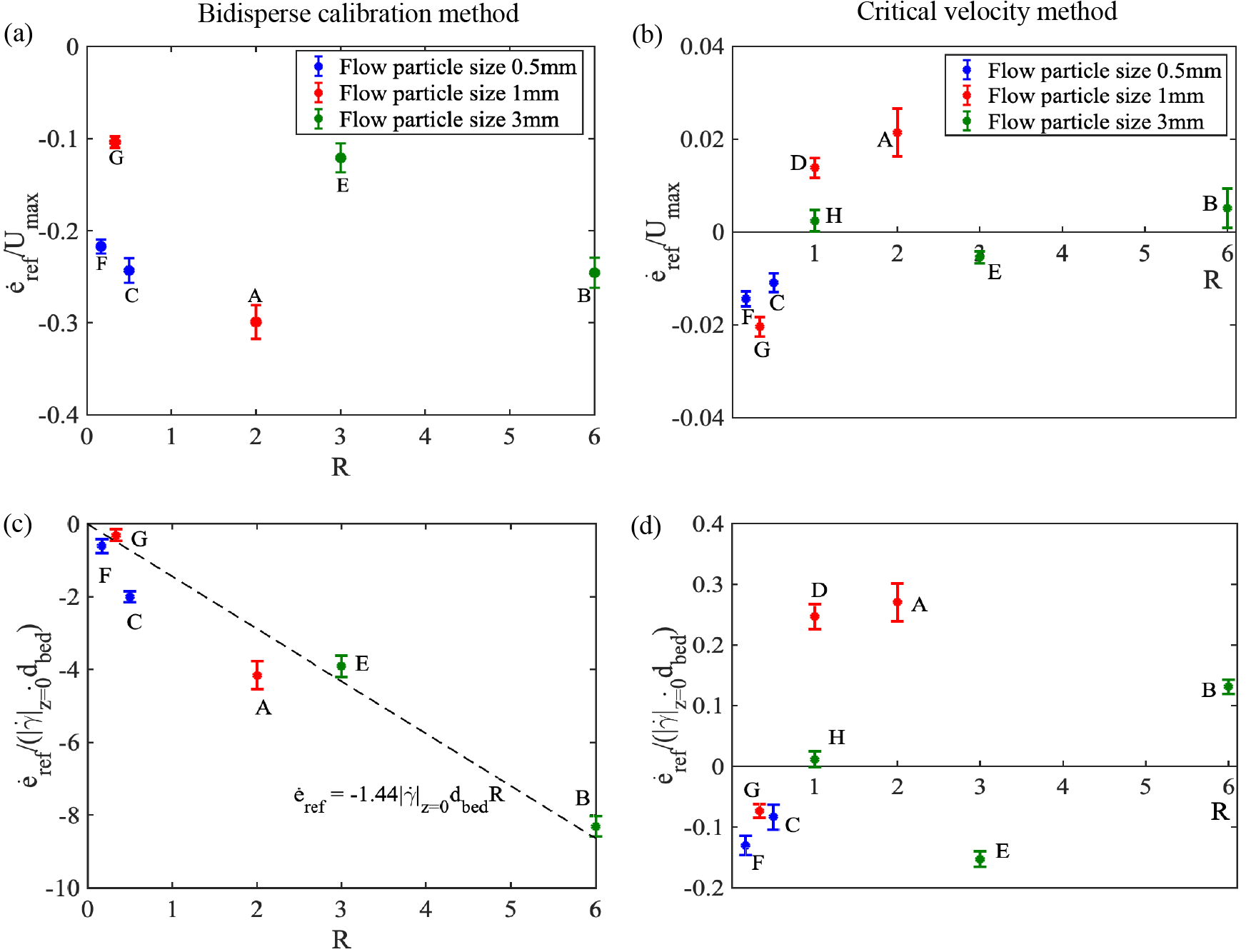}
    \caption{Comparison of normalization critical velocity and concentration based method (a),(b) Erosion rate normalized by the maximum downslope velocity. (c),(d) Erosion rate normalized by the steady flow state shear strain rate at initial position of the erodible bed and the bed particle size. Dashed line represents equation \ref{eq:best-fit-linear}.}
    \label{fig:13}
\end{figure*}

Figure~\ref{fig:13} depicts normalization of the erosion rates by the maximum velocity ((a) and (b)) or the shear strain rate at the interface and bed grain size ((c) and (d)). It is difficult to discern a trend for the critical velocity-based results, however for the bidisperse calibration method the following two scalings fit the normalised data reasonably well:

\begin{align}
    \dot{e}_{ref} &\approx -1.44|\dot\gamma|_{Z=0}d_{bed}R,\label{eq:best-fit-linear}\\
    \dot{e}_{ref} &\approx -0.25U_{max}.
\end{align}

\section{Conclusions}

The purpose of this paper was to investigate the effect of the particle size of the flowing and erodible bed material on erosion and entrainment. To evaluate the erosion experimentally, a laboratory-scale inclined flume setup was used. Several experiments were conducted with combinations of the flowing and the erodible bed particle sizes. The flow over the erodible bed was captured with dynamic X-ray radiography. Several conclusions were drawn from the results:

\begin{enumerate}
    
    \item The maximum downslope velocity increases with the flow to bed particle size ratio for a given flow particle size. 
    
    \item The erosion rates estimated from the two techniques express different contradictory conclusions. In the critical-velocity-based erosion rate estimation, the normalized erosion rate increases with the flow to bed grain size ratio. The erosion rates estimated from the bidisperse calibration approach, the normalized erosion rate decreases with increasing flow to bed grain size ratio.
    
    \item The measured erosion rate is entirely dependent on the erosion depth estimation parameter and the measuring technique. Further work is required to correlate these measurements with relevant theoretical expressions.

\end{enumerate}

\section*{Data Availability Statement}
The raw radiographs that were used to generate the results in this work are available at \cite{radiographs}. The velocity fields were produced using the freely available code PIVlab \cite{Thielicke_2014}. The bidisperse size measurements can be performed using the open source library PynamiX available at \url{https://pypi.org/project/pynamix/}.

\acknowledgments
The authors wish to acknowledge Prof Itai Einav for his valuable and constructive suggestions during this research work.

%


\bibliography{main}


\end {document}


%
%


\title{Supporting Information for "The effect of grain size on erosion and entrainment in dry granular flows"}
%
%

%
%



\authors{Eranga Dulanjalee\affil{1},
       Fran\c{c}ois Guillard\affil{1},
       James Baker\affil{1},
       Benjy Marks\affil{1}}

\affiliation{1}{School of Civil Engineering, The University of Sydney, Sydney, Australia}

%
%

%

\begin{article}

%
%

\noindent\textbf{Contents of this file}
\begin{enumerate}
\item Figures S1 to S7
\item Tables S1 to S3
\end{enumerate}

\noindent\textbf{Introduction}
In the main manuscript, a single example of data processing was provided. Here, we display the relevant information for all eight experiments performed.
\noindent\textbf{Figure S1.}
Flow height profiles of all experimental cases.
 
\noindent\textbf{Figure S2.}
Spatio-temporal distribution of the downslope velocity profiles for all cases.

\noindent\textbf{Figure S3.}
Temporally and spatially (in the $XS$ direction in each radiograph) averaged downslope velocity during steady flow state for all cases.

\noindent\textbf{Figure S4.}
Downslope velocity measured from radiography in the $XY$ direction. The flow is faster near the centre of the flume ($d=50$~mm) and slower near the side-walls at $d=0$ and $100$~mm.

\noindent\textbf{Figure S5.}
Spatiotemporal distribution of the shear strain rate profiles for all cases.

\noindent\textbf{Figure S6.}
Spatiotemporal distribution of the concentration of large particles for all cases.

\noindent\textbf{Figure S7.}
Erosion depth profiles from both the critical velocity method and the bidisperse calibration method for all cases.

\noindent\textbf{Table S1.}
Values of the shear strain rate at the initial position of the erodible bed.

\noindent\textbf{Table S2.}
Critical velocity method-based fitting parameters and goodness of fit.

\noindent\textbf{Table S3.}
Bidisperse calibration method-based fitting parameters and goodness of fit.


%
%


%
%
\bibliography{My_EndNote_Library} 
%
%
%


%
%
%
%
%

%
%
\end{article}
\clearpage

\begin{figure}
    \centering
    \includegraphics[width=1\textwidth]{Figures/supplementary/Fig_14.png}
    \caption{Flow height profiles. Letters A-H refer to the experiment cases. Zero level is the initial position of the erodible bed.}
    \label{fig:14}
 \end{figure}

\begin{figure}
    \centering
    \includegraphics[width=1\textwidth]{Figures/supplementary/Fig_15.png}
    \caption{Spatiotemporal distribution of downslope velocity profiles. Letters A-H refer to the experiment cases. Zero level is the initial position of the erodible bed. Grey region is the area with no particle flow. The dotted-grey region indicates the area with no experiment continues.}
    \label{fig:15}
 \end{figure}
 
 \begin{figure}
    \centering
    \includegraphics[width=1\textwidth]{Figures/supplementary/Fig_16.png}
    \caption{Averaged downslope velocity profiles. Letters A-H refer to experiment cases.}
    \label{fig:16}
 \end{figure}

\begin{figure}
    \centering
    \includegraphics[width=0.6\textwidth]{Figures/supplementary/Fig_17.png}
    \caption{Velocity profiles in the XY direction, indicating the non-uniformity of the flow field near the sidewalls 0 and 100 mm.}
    \label{fig:17}
 \end{figure}

\begin{figure}
    \centering
    \includegraphics[width=1\textwidth]{Figures/supplementary/Fig_18.png}
    \caption{Strain rate profiles. Letters A-H refer to the experiment cases. Zero level is the initial position of the erodible bed. Grey region is the area with no particle flow. The dotted-grey region indicates the area with no experiment continues.}
    \label{fig:18}
 \end{figure}

 
\begin{figure}
    \centering
    \includegraphics[width=0.9\textwidth]{Figures/supplementary/Fig_20.png}
    \caption{Spatio-temporal distribution of concentration of large particles. Black line is the $50\%$ concentration line of the bed grain size. Grey region is the area with no particle flow. The dotted-grey region indicates the area with no experiment continues.}
    \label{fig:20}
 \end{figure}



\begin{figure}
    \centering \includegraphics[width=0.95\textwidth]{Figures/Fig_14.png}
    \caption{Erosion depth profiles generated from critical velocity method and bidisperse calibration method.}
    \label{fig:A1}
\end{figure} 

\begin{table}
    \begin{center}
        \caption{Shear strain rate at the initial position of the erodible bed.\label{table:2}}
        \begin{tabular}{|c||l|}
            \hline Experiment & $|\dot{\gamma}|_{Z=0}~(s^{-1}) $  \\ 
            \hline
            $A$ & 4.1 \\ 
            \hline
            $B$ & 3.1 \\ 
            \hline
            $C$ & 1.1 \\ 
            \hline
            $D$ & 1.2 \\ 
            \hline
            $E$ & 1.8 \\ 
            \hline
            $F$ & 2.0 \\ 
            \hline
            $G$ & 1.8 \\ 
            \hline
            $H$ & 2.9 \\ 
            \hline
        \end{tabular}
    \end{center}
\end{table} 

\begin{table}
    \begin{center}
        \caption{Critical velocity method curve fitting parameters.\label{table:3}}
        \begin{tabular}{|c||c||c||c|}
            \hline Experiment & $Z_e^{max} (mm)$ & $t_{d}(s)$& $R-squared$ \\ 
            \hline
            $A$ & 3.9 & 8.0 & 0.01\\ 
            \hline
            $B$ & 0.9 & 5.1 & 0.1 \\ 
            \hline
            $C$ & -1.1 & 3.4 & 0.3 \\ 
            \hline
            $D$ & 3.1 & 9.4 & 0.8 \\ 
            \hline
            $E$ & -6.3 & 19.0 & 0.9 \\ 
            \hline
            $F$ & -7.4 & 16.3 & 0.6 \\ 
            \hline
            $G$ & 5.4 & 13.0 & 0.9 \\ 
            \hline
            $H$ & 1.6 & 16.9 & 0.2 \\ 
            \hline
        \end{tabular}
    \end{center}
\end{table}
 
\begin{table}
    \begin{center} 
        \caption{Bidisperse calibration method method curve fitting parameters.\label{table:4}}
        \begin{tabular}{|c||c||c||c|}
            \hline Experiment & $Z_e^{max} (mm)$ & $t_{d}(s)$& $R-squared$ \\ 
            \hline
            $A$  & -7.3 & 0.8 & 0.9 \\ 
            \hline
            $B$ & -29.8 & 2.4 & 0.9 \\ 
            \hline
            $C$ & -8.6 & 1.7 & 0.9\\ 
            \hline
            $E$ & -20.9 & 2.7 & 0.8 \\ 
            \hline
            $F$ & -6.6 & 1.8 & 0.9 \\ 
            \hline
            $G$ & -4.9 & 2.9 & 0.8 \\ 
            \hline
        \end{tabular}
    \end{center}
\end{table}